\documentclass[%
 reprint,
nofootinbib,
 amsmath,amssymb,
 aps,
]{revtex4-1}

\usepackage{graphicx}
\usepackage{dcolumn}
\usepackage{bm}
\usepackage{docs}%
\usepackage[colorlinks=true]{hyperref}

\begin{document}

\title{Singular Inflation}
\author{John D. Barrow}
\email{J.D.Barrow@damtp.cam.ac.uk}
\author{Alexander A. H. Graham}
\email{A.A.H.Graham@damtp.cam.ac.uk}
\affiliation{
Department of Applied Mathematics and Theoretical Physics\\
Centre for Mathematical Sciences, University of Cambridge, Wilberforce Road, CB3 0WA, UK 
}
\date{\today}

\begin{abstract}
We prove that a homogeneous and isotropic universe containing a scalar field with a power-law
potential, $V(\phi )=A\phi ^{n}$, with $0<n<1$ and \(A>0\) always develops a
finite-time singularity at which the Hubble rate and its first derivative
are finite, but its second derivative diverges. These are
the first examples of cosmological models with realistic matter sources that
possess weak singularities of 'sudden' type. We also show that a large
class of models with even weaker singularities exist for non-integer $n>1$.
More precisely, if $k<n<k+1$ where $k$ is a positive integer then the first divergence of
the Hubble rate occurs with its ($k+2)$th derivative. At early times these
models behave like standard large-field inflation models but they encounter
a singular end-state when inflation ends. We term this singular inflation.
\begin{description}
\item[PACS numbers] 98.80.Bp, 98.80.Cq, 98.80.Es, 98.80.Jk
\end{description}
\end{abstract}

\maketitle
\affiliation{Department of Applied Mathematics and Theoretical Physics, University of
Cambridge, Wilberforce Road, CB3 0WA, UK }

\section{\label{sec:level1}Introduction}

When can a universe possess a future singularity? \cite{BT} In the absence
of quantum effects, the best known case where this can happen, first found
by Friedmann \cite{F}, is a future 'big crunch' singularity of infinite
density. It is essentially the time reversal of the big bang singularity:
the universe expands to a maximum size at some time, contracts, and then
collapses to the future singularity. This occurs in an isotropic universe if
the cosmological constant, $\Lambda $, is negative, or if the universe is
positively curved and contains a perfect fluid with bounded pressure obeying
the strong energy condition.

A much more drastic type of future singularity is the 'big rip'. This was
first studied in Refs. \cite{starobinsky00, caldwell02, caldwell03}. It
occurs when the universe contains a fluid violating the null energy
condition. The expansion is so rapid that the expansion scale factor and all
its time-derivatives diverge in finite time. For a perfect fluid with
equation of state $p=w\rho$, linking its pressure $p$ and density $\rho $, a
'big rip' occurs when $w<-1$.

By contrast, a far weaker type of finite-time singularity is a 'sudden
singularity'. Strikingly, it does not require cosmological contraction to
occur. These singularities were first discovered in Ref. \cite{barrow86} as
a counterexample to the belief (Ref. \cite{ellis}) that compliance with the
strong and weak energy conditions and the positive pressure criterion alone
suffice to ensure that a spatially closed universe with $S^{3}$ topology has
a maximal hypersurface and recollapses. They were later discussed
systematically in Refs. \cite{barrow04, barrow04a} (see Refs. \cite%
{shtanov02, tretyakov06} for similar singularities in brane-world models).
They are characterised by the scale factor, $a(t),$ and its first derivative 
$\dot{a}(t)$ being finite as $t\rightarrow t_{s}$, but $\ddot{a}%
(t)\rightarrow \infty $ as $t\rightarrow {}t_{s}$. At these singularities $%
\rho $ is finite but $p\rightarrow +\infty $, and $\rho +3p>0$. Generalised
sudden singularities can also be constructed where the singularity occurs in
arbitrarily high derivatives of $a(t)$, with all lower derivatives finite 
\cite{barrow04a}. For further studies of these classical singularities see
Refs. \cite{barrow05, barrow09, barrow10, dabrowski05, balcerzak06,
cattoen05, nojiri05} and for discussions of quantum effects see Refs. \cite%
{fab, nojiri04}.

Sudden singularities stand in stark contrast to big rip singularities as
they are much weaker. Firstly, they satisfy all of the classical energy
conditions bar the dominant energy condition (and generalised sudden
singularities satisfy them all) \cite{lake04}. Secondly, while they are
scalar polynomial curvature singularities, they are 'weak' in the sense of
both Tipler \cite{tipler77} and Krolak \cite{krolak86} (this means they are
not \textquotedblleft crushing\textquotedblright\ singularities because an
object approaching the singularity is not crushed to zero volume). Moreover,
geodesics are actually extendible through sudden singularities \cite%
{fermandez04}, unlike for big rip and big crunch singularities\footnote{%
Note in this paper the term 'singularity' is used somewhat loosely to denote
a point where some, in principle, observable quantity (such as $H$ or its
higher order derivatives) becomes unbounded at finite time. It is not
necessarily accompanied by geodesic incompleteness, and thereby the
Hawking-Penrose singularity theorems do not apply in this case. The somewhat
informal terminology is common in the literature in this subject; a better
term for these very weak singularities might be singular events.}, and this
behaviour is stable \cite{barrow13}. For other types of cosmological
singularities discussed in the literature see Refs. \cite{gorini04,
cannata09, fermandez10, appleby10, myrzakul14, rendall05}.

Unfortunately, unlike the known examples of a big rip singularity, we do not
have a simple physically well-motivated matter model which produces a sudden
(or generalised sudden) singularity. Most studies of these singularities
postulate a form of the scale factor with the desired singularity and then
use the Einstein equations to find energy-condition compliant $\rho $ and $p$
behaviours which source this solution -- although a solution of this sort is
general in the function-counting sense \cite{barrow10}.

Here, we will construct for the first time a large family of finite-time
cosmological singularities for a canonical scalar field with a simple,
power-law potential. We show that the formation of these singularities is a
general feature of these simple matter models. They provide the first
examples of cosmological models with realistic matter which evolve towards
weak, finite-time singularities. They also have the interesting feature
that, for\ appropriate initial conditions, they can describe large-field
inflation in the early universe. The qualitative difference to standard
inflation models is that when inflation ends they evolve to a singular state
in finite time.

\section{Scalar Field Cosmologies}

We consider a spatially-flat Friedmann-Lema\^{\i}tre-Robertson-Walker (FLRW)
universe, with scale factor $a(t)$ and Hubble parameter $H(t)=\dot{a}%
(t)/a(t) $. We assume the universe contains only a scalar field, $\phi (t)$,
with self-interaction potential $V(\phi )$. The Einstein and scalar field
equations of motion are (in units where $c=8\pi {}G=1$) 
\begin{align}
& 3H^{2}=\frac{1}{2}\dot{\phi}^{2}+V(\phi ),  \label{2} \\
& \dot{H}=-\frac{1}{2}\dot{\phi}^{2},  \label{3} \\
& \ddot{\phi}+3H\dot{\phi}+V^{\prime }(\phi )=0,  \label{4}
\end{align}%
where $^{\prime }=d/d\phi $. We consider the case where the potential takes
a power-law form: 
\begin{equation}
V(\phi )=A\phi ^{n},  \label{5}
\end{equation}%
where $A>0$ is a constant. When $n$ is a positive even integer it provides
the classic example of a large-field inflation model with a single potential
minimum. If the initial conditions are chosen so that the scalar field
starts high enough up the potential then the universe inflates as the $\phi $
field rolls slowly down the potential. When it nears the potential minimum
at $\phi =0$ inflation ends and the field oscillates about the minimum. In
particular, the case where $n=4$ is the original model of chaotic inflation 
\cite{linde83}. When $n$ is a positive odd integer the universe appears to
recollapses under the influence of the scalar field (for the $n=1$ case see
Ref. \cite{pad}). We will be interested in the case where $n>0$ and is not
an integer.

\subsection{Finite-time singularities when $0<n<1$}

We first examine the case where $0<n<1$. Eq. \eqref{4} becomes 
\begin{equation}
\ddot{\phi}=-3H\dot{\phi}-An\phi ^{n-1}.  \label{6}
\end{equation}%
At $t=0$ we choose initial conditions so that $\phi _{0}\equiv \phi (0)>0$
(cases where $\phi _{0}<0$ are not cosmologically relevant, and generally
unphysical), but the value of $\dot{\phi}_{0}\equiv \dot{\phi}(0)$ is
unconstrained. We assume that the universe is expanding initially, so $%
H_{0}\equiv H(0)>0$. It is not difficult to see how the system evolves in
time. If we start with $\dot{\phi}_{0}>0$ then both terms on the right-hand
side of Eq. \eqref{6} are negative, so in a finite time $\dot{\phi}$ becomes
negative. Hence, in finite time the scalar field starts to decrease, and
since $\dot{\phi}$ continues to decrease, because the second term on the
right-hand side of Eq. \eqref{6} increases as $\phi $ decreases, it will
reach $\phi =0$ in finite time. When this happens $\ddot{\phi}\rightarrow
-\infty $ (since $n-1<0$) as $\phi \rightarrow 0$. From Eqs. \eqref{2}-%
\eqref{3}, we see $H$ and $\dot{H}$ are both finite at this point (provided
that $\dot{\phi}<0$ is finite), but $\ddot{H}$ diverges as 
\begin{equation}
\ddot{H}=-\dot{\phi}\ddot{\phi}\rightarrow -\infty \ \mbox{ as }\ \phi
\rightarrow 0.  \label{7}
\end{equation}%
Before we demonstrate these claims rigorously we make a few points about the
nature of the singularity encountered. Firstly, note it is not a scalar
polynomial (sp) curvature singularity, as both $H$ and $\dot{H}$ are finite
at this point. For a spatially-flat FLRW spacetime the Ricci scalar, $R$,
may be written as 
\begin{equation}
R=6(2H^{2}+\dot{H}),  \label{8}
\end{equation}%
which is clearly finite as $\phi \rightarrow 0$. However, higher scalar
derivatives of the curvature (like $\partial _{a}R\partial ^{a}R$ or $\Box
{}R$) are not regular since 
\begin{equation}
\dot{R}=6(4H\dot{H}+\ddot{H})\rightarrow -\infty \ \mbox{ as }\ \phi
\rightarrow 0.  \label{9}
\end{equation}%
These are generalised sudden singularities and examples of what Ellis and
Schmidt call $C^{k}$ scalar curvature singularities \cite{ES}. We might call
them scalar differential singularities. They are the first examples of such
weak singularities in an FLRW spacetime for a simple matter model (see Ref. 
\cite{coley09} for similar singularities in tilted Bianchi spacetimes).
Regularity of the curvature ensures that they are weak in the sense of both
Tipler \cite{tipler77} and Krolak \cite{krolak86}. Furthermore, since they
are weaker than sudden singularities the spacetime is geodesically complete
and extendible at the finite-time singularities.

We can ask which of the classical energy conditions these singularities
obey. A canonical scalar field is equivalent to a perfect fluid with density 
$\rho =\frac{1}{2}\dot{\phi}^{2}+V$ and pressure $p=\frac{1}{2}\dot{\phi}%
^{2}-V$. This implies that the null energy condition ($\rho +p\geq 0$) is
always satisfied for any choice of $V(\phi )$. Moreover, if $V(\phi )\geq 0$%
, as in our case, then both the weak energy condition ($\rho \geq 0$ and $%
\rho +p\geq 0$) and the dominant energy condition ($\rho \geq 0$ and $\rho
\geq |p|$) are satisfied. The strong energy condition ($\rho +p\geq 0$ and $%
\rho +3p\geq 0$) is satisfied if $\dot{\phi}^{2}\geq {}V$; while this
initially may not hold depending on the initial conditions we choose
(indeed, if this model is to function as inflation at early times it must be
violated), it is always satisfied as $\phi \rightarrow 0$. Therefore all
classical energy conditions are satisfied near the singularity. Again, this
is the first example of this kind known for a simple matter model.

Let us now prove that the system does indeed develop a finite-time
singularity of this type. It suffices to prove that the system reaches $\phi
=0$ in finite time with $\dot{\phi}$ regular at this point and $\dot{\phi}<0$%
. To do so, we prove first that the system always reaches $\dot{\phi}=0$ in
finite time, and then show that it must subsequently reach $\phi =0$ in
finite time. Note that without loss of generality we may assume that $\dot{%
\phi}_{0}>0$, as otherwise we simply omit the first step in the argument.

To prove the first claim, note that in this region of parameter space $\ddot{%
\phi}<0$, so that $\dot{\phi}$ is strictly decreasing and so, while $\phi $
increases in this region, it cannot grow faster than linearly in time: $\phi
(t)\leq \phi _{0}+\dot{\phi}_{0}t$. In Eq. \eqref{6} this implies that 
\begin{equation}
\ddot{\phi}\leq -nA\phi ^{n-1}\leq -nA(\phi _{0}+\dot{\phi}_{0}t)^{n-1}.
\label{10}
\end{equation}%
Integrating Eq. \eqref{10} once gives an upper bound on $\dot{\phi}$: 
\begin{equation}
\dot{\phi}(t)\leq \dot{\phi}_{0}+\frac{A\phi _{0}^{n}}{\dot{\phi}_{0}}-\frac{%
A}{\dot{\phi}_{0}}(\phi _{0}+\dot{\phi}_{0}t)^{n}.  \label{11}
\end{equation}%
This implies that $\dot{\phi}=0$ is reached in finite time. This completes
the first part of the argument.

For the second part, we must show that the system will reach $\phi =0$ in
finite time. Before we do so, first note that at the point where $\dot{\phi}%
=0$ we have $\ddot{\phi}=-An\phi ^{n-1}<0$, so there exists a later time, $T$%
, at which $\dot{\phi}(T)<0$, and $\dot{\phi}$ cannot become positive again.
We will prove that $\phi =0$ is reached in finite time by showing that $\dot{%
\phi}$ is a decreasing function of time, by which we mean that $\dot{\phi}%
(t)\leq \dot{\phi}(T)$ for all $t>T$. Since $\dot{\phi}$ is negative this
clearly suffices to show $\phi =0$ is reached, as integrating this
inequality once gives $\phi (t)\leq \phi (T)+(t-T)\dot{\phi}(T)$, and so $%
\phi =0$ is reached within a time $t=\phi (T)/|\dot{\phi}(T)|$ from $t=T$.

To prove that $\dot{\phi}$ is decreasing, note that without loss of
generality we may choose $T$ so that $\ddot{\phi}(T)<0$ as well. Now if $%
\dot{\phi}$ increases in the future there must exist a time $t_{1}>T$ at
which $\dot{\phi}(t_{1})=\dot{\phi}(T)$ and $\ddot{\phi}(t_{1})>0$. However,
this is impossible since Eq. \eqref{3} implies $H$ is a decreasing function
of time and at this point Eq. \eqref{6} requires 
\begin{align}
& \ddot{\phi}(t_{1})=-3H(t_{1})\dot{\phi}(t_{1})-An\phi (t_{1})^{n-1}  \notag
\label{12} \\
& <-3H(T)\dot{\phi}(T)-An\phi (T)^{n-1}=\ddot{\phi}(T)<0,
\end{align}%
hence no such time exists and so $\dot{\phi}$ is always decreasing.

While we have proven that $\phi =0$ is always reached in a finite time,
regardless of the initial conditions, we also need to show that $\dot{\phi}$
is regular at $\phi =0$. This is easily shown as $H(t)$ is a decreasing
function of time and positive in this interval, so 
\begin{equation}
\dot{\phi}^{2}(\phi =0)\leq 2A\phi ^{n}(\dot{\phi}=0).  \label{13}
\end{equation}

Since $\phi $ is finite at the turnover we conclude that $\dot{\phi}$ is
finite at $\phi =0$, and as it is a decreasing function of time we must also
have $\dot{\phi}<0$.

\subsection{Finite-time singularities when $n>1$}

Let us now examine the case where $n>1$ and $n$ is not an integer. Here, we
also expect that $\phi =0$ should be reached in a finite time from generic
initial data. In particular, it is possible to show that $\dot{\phi}=0$ is
always reached in finite time, as from Eq. \eqref{6} we have that $\ddot{\phi%
}<-nA\phi _{0}^{n-1}$, which implies that 
\begin{equation}
\dot{\phi}(t)\leq \dot{\phi}_{0}-nA\phi _{0}^{n-1}t,  \label{14}
\end{equation}%
so $\dot{\phi}=0$ is reached within time $t_{0}=\dot{\phi}_{0}/nA\phi
_{0}^{n-1}$. Moreover, at this point $\ddot{\phi}<0,$ so $\dot{\phi}$
becomes strictly negative and can never become positive, so $\phi $ always
decreases from this point. It is not so easy to show analytically that $\phi
=0$ is reached in finite time because the key step used in the argument when 
$0<n<1$ -- that $\dot{\phi}$ is always a decreasing function -- is no longer
true when $n>1$. However, it is easy to see that the only alternative is
that $\phi \rightarrow 0$ and $\dot{\phi}\rightarrow 0$ as $t\rightarrow
\infty $. This situation is at most of measure zero, and it is easy to show
via numerical simulations of Eqs. \eqref{2}-\eqref{4} that $\phi =0$ is
indeed reached in finite time for generic initial conditions for any chosen
non-integer value of $n>1$.

When $1<n<2$ we can show that $\phi =0$ is reached in finite time by the
following argument. Firstly, without loss of generality we may assume that
for $t>t_{1}>t_{0}$ we have $\ddot{\phi}\geq 0$, otherwise $\phi $ decreases
even faster and so $\phi =0$ must be reached in finite time if it is reached
when $\ddot{\phi}\geq 0$. Now $\ddot{\phi}\geq 0$ implies that for $t>t_{1}$
we have that 
\begin{equation}
\dot{\phi}\leq \frac{-An\phi ^{n-1}}{3H}\leq \frac{-An\phi ^{n-1}}{3H(t_{1})}%
,
\end{equation}%
since $H(t)$ is a decreasing function of time. Integrating this once gives 
\begin{equation}
\phi ^{2-n}\leq {}C-\frac{An(2-n)t}{3H(t_{1})},
\end{equation}%
where $C>0$ is a constant. Clearly then, if $n<2$ this implies that $\phi =0$
is reached in finite time.

When $n\geq 2$ one can use the results of Richard \cite{richard51} to derive
conditions under which $\phi =0$ is reached in finite time. Richard studied
the existence of zeros of a class of second-order, non-linear ordinary
differential equations which include Eq. \eqref{4}. Applying his results to
Eq. \eqref{4} we find a zero exists in finite time provided the function 
\begin{equation}
\theta (t)=-a(t)^{3}\frac{d}{dt}\left[ \int_{0}^{t}a(t')^{3}dt^{\prime}%
\right] ^{-\frac{2}{n+1}}
\end{equation}%
is decreasing and $\theta \rightarrow 0$ as $t\rightarrow \infty $. Notice
that in an expanding universe $\theta (t)\geq 0$, so in our case $\theta (t)$
always approaches zero from above. In general, $\theta (t)$ will be a
decreasing function provided that $a(t)$ increases slowly enough. For
instance, for a power-law expansion, $a(t)\propto {}t^{m}$, then $\theta (t)$
is a decreasing function which tends to zero asymptotically provided that 
\begin{equation}
m<\frac{1}{3}+\frac{4}{3(n-1)}.
\end{equation}%
That is, provided the expansion rate eventually approaches that of a
kinetically-dominated solution the conditions are satisfied and $\phi =0$
is reached in finite time.

When $\phi =0$ is reached Eq. \eqref{13} still implies $\dot{\phi}<0$ is
finite. Moreover, Eq. \eqref{4} now implies that $\ddot{\phi}$ is finite at
this point, unlike in the previous case. However, higher-order derivatives
of $\phi $ diverge. For instance, differentiating Eq. \eqref{4} once gives that 
\begin{equation}
\dddot{\phi}-9H^{2}\dot{\phi}-\frac{3}{2}\dot{\phi}^{3}-3HV^{\prime }(\phi
)+V^{\prime \prime }(\phi )\dot{\phi}=0.  \label{15}
\end{equation}%
For $1<n<2,$ every term is finite as $\phi \rightarrow 0$ except the last
term which is divergent, so $\dddot{\phi}\rightarrow \infty $ as $\phi
\rightarrow 0$. This implies that the first divergence in the scale factor
occurs at fourth order in its derivatives, since 
\begin{equation}
\dddot{H}=-\ddot{\phi}^{2}-\dot{\phi}\dddot{\phi}\rightarrow \infty \ 
\mbox{
as }\ \phi \rightarrow 0.  \label{16}
\end{equation}%
This implies, for instance, that $\Box {}R$ and higher derivatives of the
curvature are divergent on approach to this singularity.

It is not difficult to generalise these conclusions to arbitrarily large
non-integer values of $n$. If $k<n<k+1$, where $k$ is a positive integer,
then as $\phi \rightarrow 0$ we have $\phi ^{(k+2)}\rightarrow
(-1)^{k+1}\infty $, with all lower derivatives of $\phi $ finite. This
implies that the first divergence of the Hubble rate occurs for the $%
(k+2)^{th}$ derivative: $H^{(k+2)}\rightarrow (-1)^{k+1}\infty $ as $\phi
\rightarrow 0$. If $n$ is an integer these singularities never occur as $%
V(\phi)$ is smooth at $\phi=0$.

\section{Discussion}

We have shown that a power-law potential given by Eq. \eqref{5} admits an
arbitrarily large family of ultra-weak generalised sudden singularities
which satisfy all the classical energy conditions and are characterised by
the divergence of a sufficiently high derivative of a curvature invariant. Polynomial curvature invariants are always finite upon approach to the singularity: the divergence always occurs in a derivative of a curvature invariant due to the non-analytic behaviour of the curvature at the singularity.
More generally, we expect that any potential $V(\phi )$ which is not smooth
at $\phi =0$ will admit singularities of a similar form.

It is important to note that these models describe inflationary cosmologies
in the same way as large-field inflation models: we simply choose our
initial conditions, $\phi _{0}$ and $\dot{\phi}_{0}$, so that the system
starts high enough up the potential and is potential-dominated. Inflation
then occurs while the field violates the strong energy condition, but it
eventually ends as $\phi {}\rightarrow 0$ and the system enters the
reheating phase. The only difference is when the system reaches $\phi =0$
deep in the reheating phase (using the slow-roll approximation this is
reached in time $t\approx \frac{2}{n(4-n)}\sqrt{\frac{3}{A}}\frac{\phi
_{0}^{2-n/2}}{M_{p}}$). The difference is that reheating presumably proceeds
differently. Since predictions for the power spectrum are insensitive to the
behaviour at reheating, these models will give the same predictions for CMB
observables as large-field inflation models. Inflation with fractional
potentials has been studied before \cite{harigaya13}, and can even be
motivated from string theory \cite{silverstein08}, although their singular
behaviour as $\phi {}\rightarrow 0$ had not been recognised. Note though that the arguments used to derive fractional potentials from string theoretic constructions are only valid for large field values in general. It is also
worth noting that monomial potentials with $n<2$ given a better fit to
current CMB data than large integer values do \cite{planck}.

Let us briefly comment on the case when $n<0$ and $n$ is fractional. These
potentials are more relevant as candidates for late-time quintessence than
inflation. In this case, there is little difference between when $n$ is
fractional and when it is an integer. One can easily show that if we start
from initial data where $\phi _{0}>0$ then, generically, $\dot{\phi}$
becomes positive in finite time, and always remains positive from this point
onwards. This is because in this case $\ddot{\phi}>0$ for non-zero $\phi $
at $\dot{\phi}=0$, so if $\phi $ is initially increasing it continues to do
so forever. Even if we start from initial data where $\dot{\phi}_{0}<0$ then 
$\ddot{\phi}>0$ and so, by similar arguments to section II, $\dot{\phi}$
becomes positive in finite time. The scalar field therefore always increases
at late times, and the evolution is non-singular towards the future.

To conclude, we have demonstrated the existence of a large family of new
ultra-weak generalised sudden singularities in a spatially-flat FLRW
universe with a scalar field possessing a simple power-law potential. They
are the first examples of a spacetime possessing such weak singularities for
a simple and realistic matter model. Moreover, their formation is completely
generic in FLRW models (indeed, when $0<n<1$ they form from any isotropic
and homogeneous initial data).

The singularities discovered here are generalised sudden singularities: the
divergence in the scale factor occurs at no lower order than the third
derivative, $\dddot{a}$, (as in Ref. \cite{barrow04a}). Note that while we
can find potentials leading to these very weak singularities, and find
potentials in the same family which admit big crunch singularities, it is
not so easy to construct scalar-field models which admit sudden
singularities with divergence in $\ddot{a}$. It is not difficult to see why.
At such a singularity we would require $H$ to be finite, while $\dot{H}$
diverges. This implies $\dot{\phi}\rightarrow \infty $ on approach to the
singularity, and so $V(\phi )$ must diverge in such a way as to cancel
precisely the divergence of $\dot{\phi}$. Such potentials will likely be
unphysical, and if they do admit sudden singularities then the sensitive
cancellation required will likely render them a set of measure zero. Sudden
finite-time singularities with $\ddot{a}\rightarrow \infty $ are in some
sense a borderline case between singularities which are definitely
inextendible (big crunch or rip) and those which are not, so we could view
our results as indicating that they may not exist in general for
'reasonable' forms of matter, but their generalised forms with $%
d^{n}a/dt^{n}\rightarrow \infty $ for $n\geq 3$ are common.

Although we restricted to spatially flat universes our arguments are
unaffected if the universe is open. We also expect finite-time singularities
to form in a closed universe provided they form before the universe reaches
the expansion maximum, as in Ref. \cite{barrow86}. However, a massive scalar
field ($n=2$) in a closed FLRW universe can have quite intricate fractal
behaviour that allows a subset of initial data to avoid a strong curvature
collapse singularity \cite{page}, so we might expect analogously complicated
behaviour for non-integer potentials.

Note that we expect these singularities to be classically stable. This is
because the formation of these singularities is stable within the family of
FRLW spacetimes, so they can only be destabilised by the formation of large
anisotropies or inhomogeneities as $\phi\rightarrow0$. One can show they
also generically form in the anisotropic case, and since sudden
singularities are known to be classically stable \cite{barrow09} we expect
these even weaker singularities to be likewise stable. Indeed, as sudden
singularities have been shown to be stable \cite{barrow10} in the sense that
the homogeneous solution is the leading order solution of the field
equations containing nine independently arbitrary functions in the
neighbourhood of the singularity we would expect these finite-time
scalar-field singularities to be stable in this sense (however, the general
scalar-field solution only contains six arbitrary spatial functions on a
time slice). They are also probably even stable to quantum corrections,
since the curvature on approach to the singularity is finite and we can
construct singularities in which arbitrarily high derivatives of the
curvature are finite \cite{fab, nojiri04}. This is quite unlike most other
examples of exotic singularities known.

A pressing question raised by this paper is what is the future evolution of
these spacetimes beyond the point $\phi =0$? Since all polynomial curvature invariants
are finite, the spacetime is extendible past this point. Notice that as $%
\phi\rightarrow0$ (with $\dot{\phi}<0$) then the scale factor approaches
that of a universe dominated by a stiff fluid: $a(t)\propto{}t^{\frac{1}{3}}$%
. The precise form of $\phi(t)$ on approach to the singular point is in
general quite complicated and depends upon the precise form of $V(\phi)$.
The problem with evolving the system beyond $\phi=0$ is that in some cases
(for instance $n=\frac{1}{2}$) the matter model breaks down beyond the point 
$\phi =0$, since the naive evolution would push $\phi $ to become strictly
negative, which in general would lead to the expansion rate becoming
complex. Note though that this is only the case for some choices of $n$, and
there are many choices (e.g. $n=\frac{1}{3}$) for which $V(\phi)$ is always
real. In this case numerical evidence suggests that if $V(\phi)$ is
negative-definite for $\phi<0$ the spacetime collapses to a big crunch
singularity, while if $V(\phi)$ is positive-definite for $\phi<0$ no such
collapse occurs. Instead the universe goes through $\phi=0$ an infinite
number of times. We hope to return to this issue in future work. Some ideas
for how to approach this problem using distributional quantities are
discussed in Refs. \cite{kam1, kam2}.


A further question is how this model of inflation could be distinguished
from a similar large-field model? Also, would these singularities reported
here survive the detailed physics of reheating, which usually involves
coupling the inflaton to one or more particles? This is of course impossible
to answer without a specific model of reheating, but there is no reason to
think reheating would alter them. One way to see this is that reheating can
usually be modeled by adding a friction-like term, $\Gamma \dot{\phi}$, into
Eq. \eqref{4}, and it is not difficult to see this should not generally
affect the ability of the field to reach $\phi =0$ \cite{kofman94}. The
prospects of detecting an observable signature of this behaviour face the
same challenges as connecting the details of reheating to CMB observations.

\section*{Acknowledgements}

A.A.H.G. and J.D.B. are supported by the STFC. We thank Rahul Jha, Mary
Graham, Stephen Siklos and Alexei Starobinsky for helpful discussions.


\end{document}